# NUMERICAL ANALYSIS OF PARASITIC CROSSING COMPENSATION WITH WIRES IN DAΦNE[*]

A. Valishev, FNAL, Batavia, IL 60510, USA, D. Shatilov, BINP, Novosibirsk, Russia
C. Milardi, M. Zobov, INFN/LNF, Frascati, Italy

*Abstract*

Current bearing wire compensators were successfully used in the 2005-2006 run of the DAΦNE collider to mitigate the detrimental effects of parasitic beam-beam interactions. A marked improvement of the positron beam lifetime was observed in machine operation with the KLOE detector. In view of the possible application of wire beam-beam compensators for the High Luminosity LHC upgrade, we revisit the DAΦNE experiments. We use an improved model of the accelerator with the goal to validate the modern simulation tools and provide valuable input for the LHC upgrade project.

## INTRODUCTION

The long-range (also referred to as the parasitic) beam-beam interactions in colliders occur when the two particle beams moving in a common vacuum chamber and separated transversely, interact via the electromagnetic field. Such interactions may be a significant factor limiting the performance of multi-bunch particle colliders: for example, they were shown to impact the luminosity lifetime and increase particle losses during the Tevatron collider Run II [1]. At the Tevatron, the beams collided head-on in two high-luminosity Interaction regions (IR), and were separated by means of electrostatic separators in the rest of the machine, where each bunch experienced 70 parasitic interactions with the separation ranging from 6 to 10 of the beam σ. It was shown that four collisions at the smallest separation of 6 σ were responsible for the dramatic degradation of lifetime of both beams [2]. At the LHC, the beams collide at an angle in the experimental IRs and move in separate vacuum chambers in the arcs. Still, the number of parasitic crossings in the common sections is up to 120 with 25 ns bunch spacing at the nominal transverse separation of 9.5 σ [3]. Experiments during the LHC Run 1 have shown that with ½ the nominal number of bunches (bunch spacing of 50 ns), the onset of high losses is at the transverse separation of ≈5 σ for nominal bunch intensity [4]. The HL-LHC upgrade demands a two-fold increase in the total beam current [5], which leads to a significant enhancement of the long-range beam-beam effects [6]. As a consequence, the transverse separation of the two beams, and hence the crossing angle, has to be increased (to 12.5 σ in the baseline scenario), which in turn leads to several undesired effects: the geometric loss of luminosity, increased pile-up density, and the demand for large-aperture final focus magnets.

Compensator devices in the form of current-bearing wires (also referred to as the Beam-Beam Long-Range Compensators, BBLRC) were proposed as a way to mitigate the long-range beam-beam effects [7], and since were extensively studied both theoretically [8-10] and experimentally [11,12]. A remarkable demonstration of the effectiveness of BBLRC for improvement of collider performance was achieved during the 2005-2006 operation of DAΦNE at INFN/LNF (Frascati, Italy). The application of BBLRC devices during the KLOE run resulted in approximately 50% improvement of the average luminosity integration rate [13].

Numerical simulations of beam-beam effects with the weak-strong particle tracking code Lifetrac [14] guided the design of DAΦNE beam-beam compensation. Over the past decade, the code functionality has been considerably expanded. The most important additions include the implementation of Frequency Map Analysis (FMA) [15] and the ability to perform tracking in detailed machine lattices. The goals of the present work are to revisit the results of DAΦNE beam-beam compensation experiment using the modern computing tools, and demonstrate the predictive power of Lifetrac simulations with BBLRC for future applications.

## EXPERIMENTAL DATA

We compiled a comprehensive set of machine and beam parameters during the 2005-2006 run with the KLOE detector (see Tab. 1). The collider performance data relevant to the BBLRC experiments is presented in Figs. 1, 2 showing the time dependence of electron and positron intensities, luminosity, and the beam-beam related portion of positron losses for the cases of BBLRC turned off and on, respectively. The beam-beam related loss rate $\dot{N}_{BB}$ was derived from the total loss rate $\dot{N}$ according to the following consideration:

$$\dot{N} = \dot{N}_{Lum} + \dot{N}_T + \dot{N}_{BB}$$

where $\dot{N}_{Lum} = \sigma_{tot}L$ is the luminous loss rate ($\sigma_{tot}$ = 0.048 barn), $\dot{N}_T$ is the Touschek loss rate, and the losses due to scattering on the residual gas are vanishingly small. The typical positron loss rate in the experimental runs was about $2\times10^9$ s$^{-1}$ (corresponding beam intensity lifetime τ≈$10^3$ s), and was dominated by the Touschek and beam-beam effects with luminous losses being a relatively minor contribution at about $5\times10^6$ s$^{-1}$. The highlighted time intervals in Figs. 1, 2 are the data samples used for comparison with the beam-beam simulation as they represent stable beams in weak-strong mode. The beam-beam induced loss rate in this mode was $1.1\times10^9$ s$^{-1}$ ($\tau_{e+}=1.2\times10^3\pm175$ s) without BBLRC, and $0.55\times10^9$ s$^{-1}$ ($\tau_{e+}=2.0\times10^3\pm360$ s) with BBLRC. Clearly, the application of compensating wires resulted in a two-fold improvement of the beam-beam induced losses for identical beam parameters.

---

[*] Research supported by DOE via the US-LARP program and by EU FP7 HiLumi LHC - Grant Agreement 284404.

Table 1: DAΦNE machine and beam parameters during 2005-2006 operation with KLOE detector (March, 2006).

| Parameter | Value |
| --- | --- |
| Number of bunches | 105 |
| Bunch spacing | 2.7 ns |
| Full horizontal crossing angle | 29 mrad |
| Number of electrons / bunch | $3\times10^{10}$ |
| Number of positrons / bunch | $1\times10^{10}$ |
| Electron emittance, r.m.s. (x,y) | 0.4, 0.0056 μm |
| Positron emittance, r.m.s. (x,y) | 0.4, 0.0012 μm |
| Momentum spread, r.m.s. | $4\times10^{-4}$ |
| Bunch length, r.m.s. (e$^-$, e$^+$) | 3.0, 1.1 cm |
| Electron betatron tunes (x,y) | 0.084, 0.157 |
| Positron betatron tunes (x,y) | 0.111, 0.191 |
| Damping decrements (x,y,z) | $(9, 9, 18.1)\times10^{-6}$ |
| Beta-function at IP (x,y) | 168, 1.8 cm |
| Beam-beam parameter, e$^+$ (x,y) | 0.032, 0.029 |

## SIMULATION MODEL

The present version of Lifetrac allows importing the machine optics data for both the weak and the strong beams from MAD-X [16] files. The coordinates of particles in the weak beam are tracked through the accelerator lattice element-by-element, thus allowing for a complete treatment of the lattice nonlinearities, imperfections, strong coupling in the IR due to the detector solenoid, and chromaticity. This is a considerable improvement over the studies performed in 2004-2005, when the machine arcs were represented by linear maps.

An important attribute of Lifetrac is the capability to simulate the macroscopic measurable quantities, which allows for a direct comparison with experiment. For the case of electron-positron colliders, the code computes the equilibrium distribution for weak beam, and evaluates the specific luminosity and beam lifetime provided that the machine aperture is known [14]. We extracted the scraper positions during the experimental runs and incorporated the corresponding aperture restrictions in the modelling. The simulations were performed for the *weak* positron beam, as for the one the most affected by long-range beam-beam interactions. Parameters of the beams in simulation were as listed in Tab. 1.

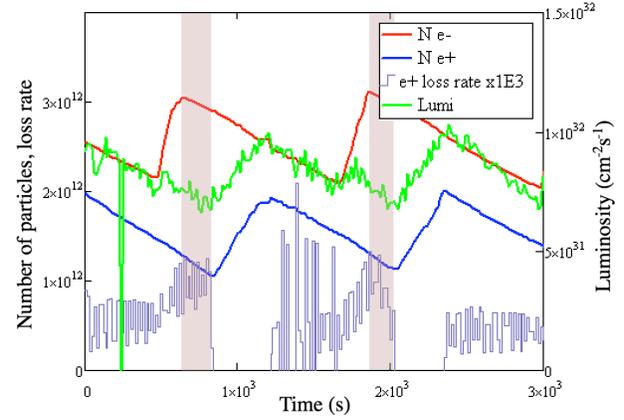

Figure 1: Evolution of beam intensity, beam-beam positron losses ($\times 10^3$ s$^{-1}$) and luminosity without beam-beam compensation.

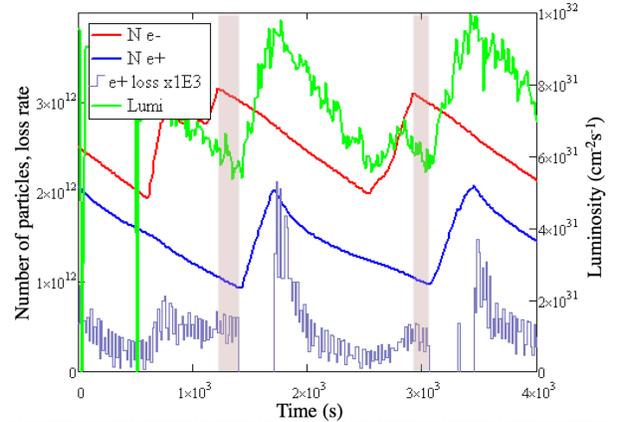

Figure 2: Evolution of beam intensity, beam-beam positron losses ($\times 10^3$ s$^{-1}$) and luminosity with beam-beam compensation.

## RESULTS

Figures 3, 4 show the FMA plots for the cases of BBLRC off and on, respectively. The area of complete overlapping of resonant islands, shown in red, represents the boundary of stable motion, the dynamical aperture (DA). It is obvious that the application of BBLRC improves the horizontal DA by some 1 σ – from 8 to 9-9.5. The effect is seen for both on- and off-momentum particles. The apparent increase of DA should result in an improvement of the injection efficiency (the beams are injected into DAΦNE in the horizontal plane), and better beam lifetime.

The results of beam density distribution calculations are presented in Figs. 5 and 6. The horizontal scrapers were placed at the distance of 9 σ, as reported by the control system read-backs, and the vertical aperture was unrestricted, and corresponded to 100 σ beam. These plots show two remarkable features: a) the size of the beam core is the same for BBLRC on and off, and corresponds to a specific luminosity of $2.2\times10^{13}$ cm$^{-2}$, which is in excellent agreement with the actual luminosity in experiment; and b) the case with wire compensators on exhibits mush smaller tail formation. The number of particles transported to large horizontal amplitudes and reaching the horizontal scrapers is smaller, which reflects in a better beam lifetime.

The results of lifetime calculations are presented in Fig. 7. We performed the calculations for different values of the limiting horizontal aperture in order to demonstrate the independence of the result on the position of the boundary. The application of BBLRC shows a marked improvement of the beam lifetime – from $2 \div 3 \times 10^3$ s without compensation to $\sim 10^5$ s with compensation in a wide range of horizontal apertures. The difference is diminishing at about 8 σ, where the tail populations for both cases become somewhat equal. It is worth noting that the lifetime calculation with the code has the accuracy of about 50%, and thus the obtained results are in a remarkable agreement with experiment.

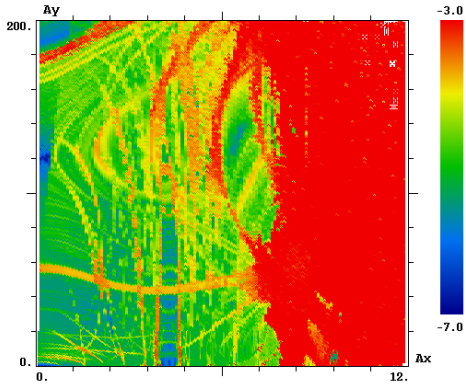

Figure 3: FMA in the space of betatron amplitudes (in units of beam σ) with BBLRC off. Colour depicts the tune jitter in logarithmic scale over the length of tracking ($2^{13}$ turns).

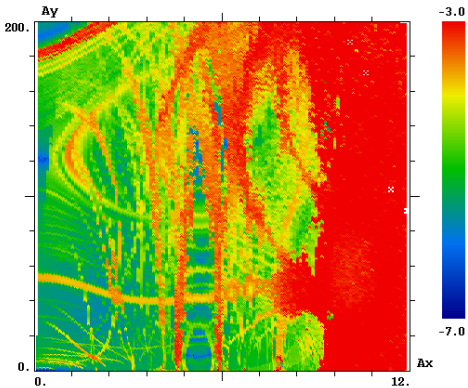

Figure 4: FMA in the space of betatron amplitudes (in units of beam σ) with BBLRC on.

## CONCLUSIONS

The experimental data from DAΦNE beam-beam compensation experiment shows a two-fold improvement of the positron beam lifetime due to the compensation at constant specific luminosity. The numerical simulations of beam-beam effects with weak-strong code Lifetrac were used to design the compensation scheme. Modeling with an improved version of the code taking full account of the machine features is in good quantitative agreement with experiment, and reproduce the macroscopic collider performance parameters, such as the specific luminosity and beam lifetime. The achieved level of precision allows to make quality predictions of beam-beam performance of future machines, such as the HL-LHC.

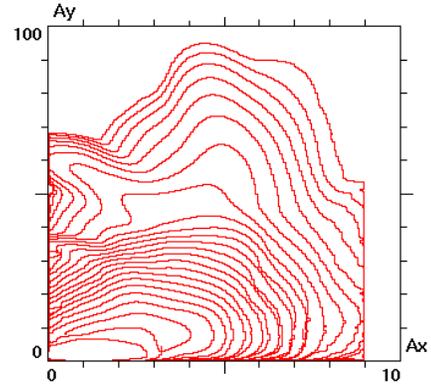

Figure 5: Lines of constant density in the space of betatron amplitudes (in units of beam σ), BBLRC off.

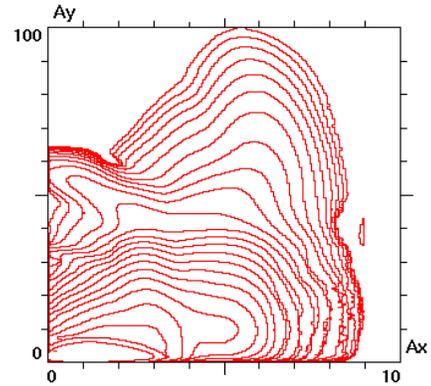

Figure 6: Lines of constant density in the space of betatron amplitudes (in units of beam σ), BBLRC on.

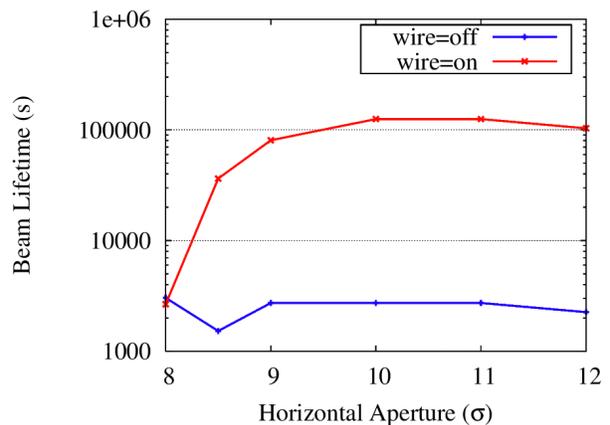

Figure 7: Simulated positron beam lifetime as a function of horizontal aperture restriction.

## ACKNOWLEDGMENT

The authors D.S. and A.V. would like to thank the DAΦNE team for hospitality during their visits.